\documentclass[aps,pra,preprint,superscriptaddress,showkeywords]{revtex4}

\usepackage{amsmath,amssymb,bm}
\usepackage{calrsfs}
\DeclareMathAlphabet{\pazocal}{OMS}{zplm}{m}{n}
\usepackage {graphicx}

\usepackage{color} 				
\usepackage[normalem]{ulem} 			

\begin{document}



\title{SU(3) symmetry in theory of a weakly interacting gas of spin-1 atoms with Bose-Einstein condensate}



\author{A.S. Peletminskii}
\email[]{aspelet@kipt.kharkov.ua}
\affiliation{Akhiezer Institute for Theoretical Physics, National Science Center
	"Kharkiv Institute of Physics and Technology", Kharkiv 61108, Ukraine \\ }
\affiliation{{V.N. Karazin Kharkiv National University, Kharkiv 61022, Ukraine}}

\author{S.V. Peletminskii}
\affiliation{Akhiezer Institute for Theoretical Physics, National Science Center
"Kharkiv Institute of Physics and Technology", Kharkiv 61108, Ukraine \\ }

\author{Yu.V. Slyusarenko}
\affiliation{Akhiezer Institute for Theoretical Physics, National Science Center
"Kharkiv Institute of Physics and Technology", Kharkiv 61108, Ukraine \\ }
\affiliation{{V.N. Karazin Kharkiv National University, Kharkiv 61022, Ukraine}}



\date{\today}

\begin{abstract}
We study a many-body system of interacting spin-1 particles in the context of homogeneous gases of ultracold atoms. In general, its description requires eight parameters among which there are three components of magnetization and five parameters associated with quadrupole degrees of freedom. Based on the symmetry considerations, we construct a many-body interaction Hamiltonian that includes eight generators of the SU(3) group related to the above description parameters. The SU(3) symmetric Hamiltonian is applied to study the ferromagnetic and quadrupolar phases of a homogeneous weakly interacting gas of spin-1 atoms with Bose-Einstein condensate. It is shown how the quadrupole degrees of freedom entering the Hamiltonian modify the ground state and single-particle excitation spectra in comparison with those obtained from the Hamiltonian bilinear in spin operators and not including quadrupole degrees of freedom. We discuss the issue of taking into account the local character of interaction to obtain the correct spectra of single-particle excitations.

\vspace{2cm}
\end{abstract}


\maketitle


\section{Introduction}

The magnetic properties of spin-1/2 crystalline systems are well understood and presented in the literature \cite{ABP,GM}. If one ignores the relativistic effects associated with the interaction of the electron magnetic moments, then the description is based on the Heisenberg model Hamiltonian and depending on the sign of the exchange integral, the system exhibits the ferromagnetic or antiferromagnetic ordering. 
This Hamiltonian is expressed in terms of the Pauli matrices representing the generators of the SU(2) group. It takes into account purely quantum (exchange) effects originating from the Fermi-Dirac statistics for electrons.

The interaction in high-spin ($S>1/2$) crystalline systems has a more complicated character that goes beyond the usual Heisenberg model and  their phase diagram exhibits a more rich structure \cite{Nagaev}. In particular, for spin-1 systems with bilinear and biquadratic exchange interactions,
the exotic orderings, such as nematic \cite{Nagaev,Matveev,Andreev,Papa} and semi-ordered \cite{Papa} phases, may exist along with the traditional ferromagnetic and antiferromagnetic phases. Moreover, the non-Heisenberg  structure of the spin-spin interaction affects even the traditional phases. In recent years, there have been intensive studies of unconventional orderings in magnets with $S\geqslant 1$ \cite{Harada,Toth,BH,NC,Fridman,Kosm,Kolezhuk,Kov1}.

Nowadays, the interest in high-spin systems is attracted by the studies of ultracold atomic gases providing remarkable opportunities to examine and model various effects and phenomena in quantum many-body systems in a well-controlled manner. In particular, quantum gases loaded in an optical lattice represent
an artificial but effective simulator of magnetic phenomena in crystalline systems \cite{Imambekov,Rodriguez,Sotnikov1,Sotnikov2}. The first theoretical investigations of magnetic phases and corresponding excitations in dilute homogeneous (or trapped) Bose gases with condensate \cite{Ohmi,APS,Ho} were stimulated by experiments on optical trapping a condensate of ${}^{23}$Na spin-1 atoms \cite{Stamper}. Comprehensive reviews of the so-called spinor Bose gases, including those of spin-2 and spin-3 atoms, were presented in Refs. \cite{Ueda,Kurn,Yukalov}. In most of these studies, the interaction between the internal degrees of freedom was parametrized by the scattering length and the corresponding Hamiltonian was considered as a bilinear form in spin operators, like in the usual Heisenberg model.
However, as mentioned, such a form of interaction energy is not sufficient to describe properly the magnetic properties of high-spin systems.

In this paper, proceeding solely from the symmetry considerations, we propose a general recipe for obtaining the many-body Hamiltonian describing a system of interacting spin-$S$ atoms. 
Such systems are characterized by $(2S+1)^{2}-1$ parameters among which there are three components of magnetization vector and the rest can be treated as the multipole degrees of freedom. These additional parameters are induced in a many-body system by the spin of the structural constituents of matter and appear at the macroscopic level.  As the most intriguing case with a view to physics of ultracold Bose gases, we present a detailed study of the many-body Hamiltonians describing the interaction effects in the system of spin-1 atoms.  
Then we apply the SU(3) symmetric Hamiltonian with quadrupole degrees of freedom to examine the ferromagnetic and quadrupolar phases of a weakly interacting Bose gas with condensate in a magnetic field. It is shown that the quadrupole degrees of fredom modify the ground state properties and single-particle excitation spectra of the system.

\section{Formulation of the problem and many-body interaction Hamiltonian of internal degrees of freedom}

Consider a many-body system of spin-$S$ particles whose reduced description is performed in terms of the single-particle density matrix $f_{\alpha\beta}({\bf p})= {\rm Tr}\,\varrho a^{\dagger}_{{\bf p}\beta}a_{{\bf p}\alpha}$, where $\varrho$ can be either an equilibrium or non-equilibrium statistical operator, $a^{\dagger}_{{\bf p}\alpha}$, $a_{{\bf p}\alpha}$ are the creation and annihilation operators with index $\alpha$ running $2S+1$ values. Depending on the spin value, these operators meet the following bosonic commutation (integer spin) or fermionic anticommutation (half-integer spin) relations:
\begin{gather}
[a_{{\bf p}\alpha},a^{\dagger}_{{\bf p}'\alpha'}]_{B}=\delta_{{\bf p}{\bf p}'}\delta_{\alpha\alpha'}, \quad [a_{{\bf p}\alpha},a_{{\bf p}'\alpha'}]_{B}=0, \label{eq:2.1} \\
\{a_{{\bf p}\alpha},a^{\dagger}_{{\bf p}'\alpha'}\}_{F}=\delta_{{\bf p}{\bf p}'}\delta_{\alpha\alpha'}, \quad \{a_{{\bf p }\alpha},a_{{\bf p}'\alpha'}\}_{F}=0. \label{eq:2.2}
\end{gather}
Since below we study a homogeneous interacting Bose gas, we use the momentum ${\bf p}$ to specify the individual state of a particle. For the lattice models, one should consider a lattice site index instead of ${\bf p}$.

In the case of spin-1/2 system, the density matrix $f_{\alpha\beta}({\bf p})$, being a square matrix of the second order, can be written as a linear combination of the Pauli matrices $\sigma^{i}_{\alpha\beta}$ and unit matrix $I$ which form a basis for the vector space of $2\times 2$ matrices.
The scalar part of the single-particle density matrix in such a decomposition defines the density of the system, whereas its vectorial part specifies three components of the magnetization vector. The latter quantity is induced by the atomic spin.
The many-body Hamiltonian of two-particle interaction includes the so-called spin-spin interaction given by
\begin{equation} \label{eq:2.3}
V={1\over 2\mathcal{V}}\sum_{{\bf p}_{1},\ldots{\bf p}_{4}}I({\bf p}_{1}-{\bf p}_{3})a^{\dagger}_{{\bf p}_{1}\alpha}a^{\dagger}_{{\bf p}_{2}\beta}S^{i}_{\alpha\gamma}S^{i}_{\beta\delta}a_{{\bf p}_{3}\gamma}a_{{\bf p}_{4}\delta}\,\delta_{{\bf p}_{1}+{\bf p}_{2},\,{\bf p}_{3}+{\bf p}_{4}}, \quad S^{i}_{\alpha\beta}={1\over 2 }\sigma^{i}_{\alpha\beta},
\end{equation}
where $a^{\dagger}_{{\bf p}\alpha}$, $a_{{\bf p}\alpha}$ satisfy the permutation relations given by Eqs.~(\ref{eq:2.2}) and $I({\bf p}_{1}-{\bf p}_{3})$ denotes the Fourier transform of exchange interaction energy. Here and below, the summation over the repeated indices related to internal symmetry is assumed. The above Hamiltonian commutes with the spin operator of a many-body system,
$$
S^{i}=\sum_{\bf p}a^{\dagger}_{{\bf p}\alpha}S^{i}_{\alpha\beta}a_{{\bf p}\beta}
$$
and, consequently, the latter represents the conserved quantity or the integral of motion. It is related to  magnetization vector $M^{i}$ by $M^{i}=g\mu_{B}S^{i}$, where $g$ is the Land\'{e} hyperfine $g$-factor and $\mu_{B}=e\hbar/2m_{e}$ is the Bohr magneton. Note that three components of an atomic spin (microscopic characteristic) generate the same number of macroscopic parameters necessary to describe a many-body system of spin-1/2 particles. The Hamiltonian given by Eq.~(\ref{eq:2.3}) can be applied to describe a gas of spin-1/2 atoms or interacting electron gas embedded in a solid state system. Note that in the lattice models, the creation and annihilation operators carry lattice site index instead of momentum ${\bf p}$.

Now we address the description of a many-body system of spin-1 particles. In this case, the single-particle density matrix, being a reduced description parameter, can be written as a linear combination of the unit $3\times 3$ matrix $I_{\alpha\beta}$ and the Gell-Mann linearly independent traceless Hermitian matrices $\lambda^{a}_{\alpha\beta}$ (see Appendix):
\begin{equation} \label{eq:2.4}
f_{\alpha\beta}({\bf p})=f^{0}({\bf p})I_{\alpha\beta}+f^{a}({\bf p})\lambda_{\alpha\beta}^{a}, \quad a=1,\dots,8.
\end{equation}
The scalar $f^{0}({\bf p})$ and vectorial $f^{a}({\bf p})$ coefficients are given by
\begin{equation} \label{eq:2.5}
f^{0}={1\over 3}{\rm Tr}\,f({\bf p}), \quad f^{a}({\bf p})={1\over 2}{\rm Tr}\,f({\bf p})\lambda^a.
\end{equation}
In contrast to spin-1/2 systems, we see that three components of spin are insufficient to describe the many-body states of the system. Indeed, according to Eqs.~(\ref{eq:2.4}), (\ref{eq:2.5}), the states are specified by eight independent parameters determined by generators $\lambda^{a}$ of the SU(3) group.

To clarify the physical meaning of eight parameters associated with internal symmetry, consider the realization of spin-1 operators in the vector (Cartesian) basis $|x\rangle$, $|y\rangle$, $|z\rangle$ instead of the usual canonical (irreducible) basis $|S,m\rangle$ with $S=1$ and $m=-1,0,1$. These two are related by (see e.g. \cite{Papa}):
$$
|x\rangle=\sqrt{1/2}\left(|1,1\rangle+|1,-1\rangle\right), \quad |y\rangle=-i|1,0\rangle,
\quad |z\rangle=i\sqrt{1/2}\left(|1,1\rangle-|1,-1\rangle\right).
$$
In the vector basis we have
\begin{equation} \label{eq:2.6}
\langle i|k\rangle=\delta_{ik}, \quad S^{i}|k\rangle=i\varepsilon_{ikl}|l\rangle,
\end{equation}
so that $S^{i}$ meet the usual commutation relations for spin operators,
$$
[S^{i},S^{k}]=i\varepsilon_{ikl}S^{l}.
$$
From Eq.~(\ref{eq:2.6}), one finds the matrix elements for the corresponding spin operators,
$$
\langle k|S^{i}|l\rangle\equiv (S^{i})_{kl}=-i\varepsilon_{ikl},
$$
whence
\begin{equation*} \label{eq:2.7}
S^{x}=\left(
  \begin{array}{ccc}
    0 & 0 & 0 \\ [-2ex]
    0 & 0 & -i \\ [-2ex]
    0 & i & 0 \\
  \end{array}
\right), \quad
S^{y}=\left(
  \begin{array}{ccc}
    0 & 0 & i \\[-2ex]
    0 & 0 & 0 \\[-2ex]
    -i & 0 & 0 \\
  \end{array}
\right), \quad
S^{z}=\left(
  \begin{array}{ccc}
    0 & -i & 0 \\ [-2ex]
    i & 0 & 0 \\ [-2ex]
    0 & 0 & 0 \\
  \end{array}
\right).
\end{equation*}
One can easily seen that $S^{x}=\lambda^{7}$, $S^{y}=-\lambda^{5}$, and $S^{z}=\lambda^{2}$ (see Eqs.~(\ref{eq:A1})), so that subalgebra of these matrices generates an SU(2) subgroup of SU(3) group. The remaining five Gell-Mann matrices, due to their properties given by Eq.~(\ref{eq:A6}), can be expressed in terms of the quadratic combinations of spin operators:
\begin{gather*}
\lambda^{1}=-\{S^{x},S^{y}\}, \quad \lambda^{3}=(S^{y})^{2}-(S^{x})^{2}, \quad \lambda^{4}=-\{S^{x},S^{z}\}, \nonumber \\
\lambda^{6}=-\{S^{y},S^{z}\}, \quad \lambda^{8}=\sqrt{3}(S^{z})^{2}-{2\over\sqrt{3}}\,I, \label{eq:2.8}
\end{gather*}
where $\{\ldots,\ldots\}$ denotes an anticommutator and $I$ is the unit $3\times3$ matrix. Since the traceless quadrupole matrix $Q^{ik}\equiv S^{i}S^{k}+S^{k}S^{i}-(4/3)\,\delta_{ik}$ is determined by the above five independent components,
$$
Q^{ik}=\left(
  \begin{array}{ccc}
    -\lambda^{3}-{\dfrac{\lambda^{8}}{\sqrt{3}}} & -\lambda^{1} & -\lambda^{4} \\ 
    -\lambda^{1} & \lambda^{3}-{\dfrac{\lambda^{8}}{\sqrt{3}}} & -\lambda^{6} \\
    -\lambda^{4} & -\lambda^{6} & {\dfrac{{2\lambda^{8}}}{\sqrt{3}}} \\
  \end{array}
\right),
$$
we call them  the quadrupole operators. These operators can be considered as the components of a single vector $q^{b}=(\lambda^{1},\lambda^{3},\lambda^{4},\lambda^{6},\lambda^{8})$. Therefore, if the microscopic constituents of a many-body system have a unit spin, then its macroscopic state is described by the eight parameters originating from the generators of the SU(3) group,
\begin{equation}\label{eq:2.9}
\Lambda^{a}=\sum_{\bf p}a^{\dagger}_{{\bf p}\alpha}\lambda^{a}_{\alpha\beta}a_{{\bf p}\beta}, \quad a=1,\ldots 8, \quad \alpha,\, \beta=x,\,y,\,z,
\end{equation}
which can be split into the spin and quadrupole operators,
\begin{equation} \label{eq:2.10}
S^{i}=\sum_{\bf p}a^{\dagger}_{{\bf p}\alpha}S^{i}_{\alpha\beta}a_{{\bf p}\beta} \quad (i=x,\,y,\,z), \quad Q^{b}=\sum_{\bf p}a^{\dagger}_{{\bf p}\alpha}q^{b}_{\alpha\beta}a_{{\bf p}\beta} \quad (b=1,\,3,\,4,\,6,\,8).
\end{equation}

In the second quantization method a general two-body operator (or binary operator) $A^{(2)}$ can be represented in the form (see, e.g., \cite{MethStatPhys})
\begin{equation}\label{eq:2.10'}
A^{(2)}={1\over 4}\sum_{\alpha_{1},\ldots \alpha_{4}}a^{\dagger}_{\alpha_{1}}a^{\dagger}_{\alpha_{2}}A_{\alpha_{1}\alpha_{2};\alpha_{3}\alpha_{4}} a_{\alpha_{3}}a_{\alpha_{4}}, \quad A_{\alpha_{1}\alpha_{2};\alpha_{3}\alpha_{4}}=\langle \alpha_{1},\alpha_{2}|A^{(2)}|\alpha_{3},\alpha_{4}\rangle.
\end{equation}
Therefore, one can write a two-body Hamiltonian describing the interaction of internal degrees of freedom in a many-body system of spin-1 particles in the following form:
\begin{gather}
V_{J-K}={1\over 2\mathcal{V}}\sum_{{\bf p}_{1},\ldots{\bf p}_{4}}J({\bf p}_{1}-{\bf p}_{3})a^{\dagger}_{{\bf p}_{1}\alpha}a^{\dagger}_{{\bf p}_{2}\beta}S^{i}_{\alpha \gamma}S^{i}_{\beta\delta}a_{{\bf p}_{3}\gamma}a_{{\bf p}_{4}\delta}\,\delta_{{\bf p}_{1}+{\bf p}_{2},\,{\bf p}_{3}+{\bf p}_{4}} \nonumber \\ 
+{1\over 2\mathcal{V}}\sum_{{\bf p}_{1},\ldots{\bf p}_{4}}K({\bf p}_{1}-{\bf p}_{3})a^{\dagger}_{{\bf p}_{1}\alpha}a^{\dagger}_{{\bf p}_{2}\beta}q^{b}_{\alpha \gamma}q^{b}_{\beta\delta}a_{{\bf p}_{3}\gamma}a_{{\bf p}_{4}\delta}\,\delta_{{\bf p}_{1}+{\bf p}_{2},\,{\bf p}_{3}+{\bf p}_{4}}, \label{eq:2.11} 
\end{gather}
where ${\cal V}$ is the volume of the system and $J({\bf p}_{1}-{\bf p}_{3})$, $K({\bf p}_{1}-{\bf p}_{3})$ are the Fourier transforms of the corresponding interaction energies. The above Hamiltonian has SU(2) symmetry (see Eqs.~(\ref{eq:2.12}), (\ref{eq:2.12'}) below). However, it becomes SU(3) symmetric when $J=K$,
\begin{equation}\label{eq:2.11'}
V_{J}={1\over 2\mathcal{V}}\sum_{{\bf p}_{1},\ldots{\bf p}_{4}}J({\bf p}_{1}-{\bf p}_{3})a^{\dagger}_{{\bf p}_{1}\alpha}a^{\dagger}_{{\bf p}_{2}\beta}\lambda^{a}_{\alpha \gamma}\lambda^{a}_{\beta\delta}a_{{\bf p}_{3}\gamma}a_{{\bf p}_{4}\delta}\,\delta_{{\bf p}_{1}+{\bf p}_{2},\,{\bf p}_{3}+{\bf p}_{4}}.
\end{equation}
Indeed, taking into account the commutation relations for the creation and annihilation operators as well as the properties of the structure constants $f^{abc}$ (see Eqs.~(\ref{eq:2.1}), (\ref{eq:A4})), one can show that $[V_{J},\Lambda^{a}]=0$ and, consequently, $\Lambda_{a}$ is the integral of motion. Thus, the Hamiltonian given by Eq.~(\ref{eq:2.11'}) is SU(3) invariant. This symmetry, however, is broken if the Zeeman coupling between the spin and magnetic field is taken into account. The role of SU(3) symmetry in the dynamics and relaxation of spin-1 magnets is of much current interest \cite{Kov2,Kov3,Ivanov}. 

Using Eqs.~(\ref{eq:A9}), (\ref{eq:A10}), one can show that
\begin{equation} \label{eq:2.12}
{1\over 2}q^{b}_{\alpha\gamma}q^{b}_{\beta\delta}=S^{i}_{\alpha\sigma}S^{i}_{\beta\rho}S^{k}_{\sigma\gamma}S^{k}_{\rho\delta}+{1\over 2}S^{i}_{\alpha\gamma}S^{i}_{\beta\delta}-{4\over 3}\delta_{\alpha\gamma}\delta_{\beta\delta} 
\end{equation}
or equivalently
\begin{equation} \label{eq:2.12'}
{1\over 2}\lambda^{a}_{\alpha\gamma}\lambda^{a}_{\beta\delta}=S^{i}_{\alpha\gamma}S^{i}_{\beta\delta}+S^{i}_{\alpha\sigma}S^{i}_{\beta\rho}S^{k}_{\sigma\gamma}S^{k}_{\rho\delta}-{4\over 3}\delta_{\alpha\gamma}\delta_{\beta\delta}. 
\end{equation}
Consequently, the appearance of quadrupole degrees of freedom in Eqs.~(\ref{eq:2.11}), (\ref{eq:2.11'}) is equivalent to the fact that the Hamiltonian contains both bilinear and biquadratic terms in spin operators. 

Now we address the issue of when it is necessary to take into account the quadrupole degrees of freedom. To this end, consider a collision of two bosonic atoms of spin $s=1$. For the sake of simplicity, their interaction is assumed to be parametrized by three coupling constants $g_{\pazocal{S}}$ (for a while, they are not related to $s$-wave scattering lengths) in the total spin $\pazocal{S}$ channel, $V=g_{0}P_{0}+g_{1}P_{1}+g_{2}P_{2}$. Here $P_{\pazocal{S}}$ ($P_{\pazocal{S}}P_{\pazocal{S}'}=\delta_{\pazocal{S}\pazocal{S}'}$, $P_{\pazocal{S}}^{2}=P_{\pazocal{S}}$) is the projection operator which projects the wave function of a pair of atoms into a total spin $\pazocal{S}$ state. The relation ${\bf S}_{1}\cdot{\bf S}_{2}=\sum_{\pazocal{S}=0}^{2}\lambda_{\pazocal{S}}P_{\pazocal{S}}$ with $\lambda_{\pazocal{S}}={1\over 2}[\pazocal{S}(\pazocal{S}+1)-2s(s+1)]$ gives
\begin{gather*}
{\bf S}_{1}\cdot{\bf S}_{2}=-2P_{0}-P_{1}+P_{2}, \\
({\bf S}_{1}\cdot{\bf S}_{2})^{2}=P_{2}+P_{1}+4P_{0}
\end{gather*}
Adding here the completeness condition for the projection operator, $P_{0}+P_{1}+P_{2}=1$, 
one obtains the the system of coupled equations for determining $P_{\pazocal{S}}$. Its solution reads 
\begin{gather*}
P_{0}={1\over 3}\left[({\bf S}_{1}\cdot{\bf S}_{2})^{2}-1\right], \\
P_{1}=1-{1\over 2}\left[({\bf S}_{1}\cdot{\bf S}_{2})+({\bf S}_{1}\cdot{\bf S}_{2})^{2}\right], \\
P_{2}={1\over 3}+{1\over 2}({\bf S}_{1}\cdot{\bf S}_{2})+{1\over 6}({\bf S}_{1}\cdot{\bf S}_{2})^{2}.
\end{gather*}
Now the interaction takes the form 
\begin{equation}\label{eq:2.13}
V=c_{0}+c_{1}({\bf S}_{1}\cdot{\bf S}_{2})+c_{2}({\bf S}_{1}\cdot{\bf S}_{2})^{2},
\end{equation}
where
$$
c_{0}=g_{1}+{1\over 3}\left(g_{2}-g_{0}\right), \quad  c_{1}={1\over 2}\left(g_{2}-g_{1}\right), \quad c_{2}={1\over 3}\left(g_{0}-{3g_{1}\over 2}+{g_{2}\over 2}\right).
$$
Therefore, in general case of three channel scattering with total spin $\pazocal{S}=0,1,2$, it is necessary to take into account the biquadratic term $({\bf S}_{1}\cdot{\bf S}_{2})^{2}$ in the interaction Hamiltonian. According to Eqs.~(\ref{eq:2.12}), (\ref{eq:2.12'}), this is equivalent to the fact that the Hamiltonian contains the quadrupole degrees of freedom. 

However, if the interaction is parametrized by the scattering lengths $a_{\pazocal{S}}$ corresponding to $s$-state of the relative motion, so that $g_{\pazocal{S}}=4\pi\hbar^{2}a_{\pazocal{S}}/m$ ($m$ is the mass of the atom), then the scattering with the total spin $\pazocal{S}=1$ is forbidden due to requirement for the wave function to be symmetric under exchange of two atoms. In this case, the projection operator into a state of total angular momentum $\pazocal{S}=1$ is zero, $P_{1}=0$, which leads to the relation $({\bf S}_{1}\cdot{\bf S}_{2})^{2}=2-{\bf S}_{1}\cdot{\bf S}_{2}$. The latter allows us to write Eq.~(\ref{eq:2.13}) in the form, $V=\tilde{c}_{0}+\tilde{c}_{2}{\bf S}_{1}\cdot{\bf S}_{2}$, where $\tilde{c}_{0}={1\over 3}(g_{0}+2g_{2})$ and $\tilde{c}_{2}={1\over 3}(g_{2}-g_{0})$ \cite{Ho}. Thus, the description of interacting system by the scattering lengths does not require to take into account the biquadratic terms in spin operators or quadrupole degrees of freedom. For this reason, in order to study the effects of quadrupole degrees of freedom in ultracold gases, we consider the general interaction Hamiltonians (see Eqs.~(\ref{eq:2.11}), (\ref{eq:2.11'})). In this regard, we note that in spite of the fact that the interaction effects in ultracold gases are fairly well approximated by the scattering lengths, this approximation is not so "harmless" (see Refs. \cite{JPhysB,Hazlett,Caballero,JPhysB2,Ferm-nonloc1} and the discussion below).

The structure of the SU(3) symmetric Hamiltonian can also be justified within a phenomenological quasiparticle theory, where the energy of the system is considered to be a functional of the single-particle density matrix, like in the normal Fermi-liquid theory \cite{Landau,Silin}. For a not dense system, one can restrict ourselves by the energy functional quadratic in the single-particle density matrix \cite{FNT}. In the case of spin-1 system, the phenomenological interaction amplitude, being the second variational derivative of the energy functional, has four indices associated with internal symmetry which are summed over with those of single-particle density matrices. It is clear that the interaction part of this functional should be related to the general form of the pair interaction microscopic Hamiltonian (see Eq.~(\ref{eq:2.11'})). This relation is given by statistical averaging of the latter and using the Bloch--De Dominicis (or Wick's) theorem \cite{BogBog}.
Then, decomposing the interaction amplitude in the energy functional over the complete set of three-row Gell-Mann matrices with respect to each pair of indices and requiring it to commute with all generators of the SU(3) group, one can arrive at the interaction structure in the form of Eq.~(\ref{eq:2.11'}).

Finally, note that the formalism of the SU(3) Lie algebra was earlier involved to study the spin-1 lattice models whose Hamiltonians are constructed of the group generators related to spin and quadrupole operators \cite{BH,Onufrieva1,Onufrieva2}. 
In contrast to the above studies dealing with lattice models, the proposed approach allows to treat a homogeneous (or trapped) quantum gas of spin-1 atoms. In particular, the Hamiltonian in the form of Eq.~(\ref{eq:2.11'}) is applied below to examine the ground state structure and corresponding excitations of a weakly interacting gas of spin-1 atoms taking into account the quadrupole degrees of freedom. In previous studies \cite{Ohmi,APS,Ho}, such a problem was analyzed on the basis of the Hamiltonian with bilinear spin interaction not including quadrupole operators. As for the lattice models, the Hamiltonians in the form of Eqs.~(\ref{eq:2.11}), (\ref{eq:2.11'}) can be used to derive the appropriate Bose-Habbard Hamiltonian describing the properties of an ultracold dilute gas of bosonic spin-1 atoms in optical lattice \cite{Jaksch}. Such spinor gases in optical lattices  provide a novel realization of quantum magnetic systems \cite{Imambekov,Rodriguez,Chiara}.

\section{Truncated Hamiltonian for a weakly interacting Bose gas with internal degrees of freedom}

As we noted, ultracold quantum gases provide a powerful tool to study various effects and phenomena in quantum many-body systems. Therefore, we apply the obtained SU(3) symmetric interaction Hamiltonian in the form of Eq.~(\ref{eq:2.11'}) to study the ground state structure and single-particle excitations of a weakly interacting Bose gas of spin-1 atoms with Bose-Einstein condensate. This Hamiltonian follows from a more general Hamiltonian given by Eq.~(\ref{eq:2.11}) when $J=K$. This can be achieved, e.g., by means of a Feshbach resonance. The role of SU(3) symmetry in the dynamics and relaxation of spin-1 magnets is of much current interest \cite{Kov2,Kov3,Ivanov}.

In our study, we employ the Bogoliubov model \cite{Bogoliubov} based on $c$-number treatment of creation and annihilation operators for condensate particles. Our starting point is the following Hamiltonian consisting of the kinetic energy term $H_{0}$ and the terms corresponding to the potential interaction $V_{U}$ as well as the interaction between the internal degrees of freedom $V_{J}$:
\begin{equation} \label{eq:3.1}
H=H_{0}+V_{U}+V_{J},
\end{equation}
where
\begin{gather}
H_{0}=\sum_{\bf p}a^{\dagger}_{{\bf p}\alpha}\left[\varepsilon_{\bf p}\delta_{\alpha\beta}- hS^{z}_{\alpha\beta}\right]a_{{\bf p}\beta}, \quad S^{z}_{\alpha\beta}\equiv\lambda^{2}_{\alpha\beta}, \label{eq:3.2} \\
V_{U}={1\over 2\cal{V}}\sum_{{\bf p}_{1},\ldots{\bf p}_{4}}U({\bf p}_{1}-{\bf p}_{3})a^{\dagger}_{{\bf p}_{1}\alpha}a^{\dagger}_{{\bf p}_{2}\beta}a_{{\bf p}_{3}\alpha}a_{{\bf p}_{4}\beta}\,\delta_{{\bf p}_{1}+{\bf p}_{2},\,{\bf p}_{3}+{\bf p}_{4}}, \label{eq:3.3}
\end{gather}
and $V_{J}$ is given by Eq.~(\ref{eq:2.11'}). Here $\varepsilon_{\bf p}=p^{2}/2m$ is the kinetic energy of a particle, $U({\bf p})$ is the Fourier transform for the potential interaction energy, and
$h=g\mu_{B}H$, where $g$, $\mu_{B}$, and $H$ are the Land\'{e} hyperfine factor \cite{Ueda}, the Bohr magneton, and external magnetic field directed along $z$-axis, respectively. Note that usually the interaction Hamiltonian is written in terms of the corresponding scattering lengths describing the low energy collisions of atoms at ultra low temperature \cite{Ohmi,Ho,Ueda,Kurn}. However, such parametrization of interaction does not take into account the quadrupole degrees of freedom and local character of interaction. The latter results in divergences when computing the ground state energy or chemical potential so that it is necessary to use the renormalization of the coupling constant \cite{Pethick,Stringari,JPhys2017}.  Moreover, as we see below, it may lead to an incomplete structure of the spectrum of single-particle excitations. Therefore, the interaction given by Eqs.~(\ref{eq:2.11'}), (\ref{eq:3.3}) is characterized by the corresponding functions $U({\bf p})$ and $J({\bf p})$.

Since the number of Bose condensed atoms is a macroscopic value proportional to the volume of the system ${\cal V}$, the next step, according to the Bogoliubov model \cite{Bogoliubov}, is to replace the creation and annihilation operators of condensed atoms with zero momentum by $c$-numbers $a_{0}^{\dagger}\to\sqrt{\cal V}\Psi^{*}_{\alpha}$ and $a_{0}\to\sqrt{\cal V}\Psi_{\alpha}$ in all operators of relevant physical quantities, where $\Psi_{\alpha}$ represents the condensate wave function.
This procedure has been proved to be exact in the thermodynamic limit \cite{Ginibre}. The $c$-number terms in the Hamiltonian and those that are quadratic in creation and annihilation operators allow us to define the ground state and the corresponding spectra of single-particle excitations (quasiparticles), whereas the higher order terms in creation and annihilation operators are relevant when describing the interaction effects between the quasiparticles themselves. Therefore, performing the above replacement in Eqs.~(\ref{eq:3.1})-(\ref{eq:3.3}) with $\Psi_{\alpha}$ being a variational parameter and neglecting the terms of the third and fourth order, one can obtain the Hamiltonian truncated up to quadratic terms in the creation and annihilation operators:
\begin{equation} \label{eq:3.5}
H(\Psi)\simeq H^{(0)}(\Psi)+H^{(2)}(\Psi),
\end{equation}
where
$H^{(0)}(\Psi)$ is the $c$-number part of the truncated Hamiltonian given by
\begin{equation} \label{eq:3.6}
{1\over{\cal V}}H^{(0)}(\Psi)={U(0)\over 2}(\Psi^{*}\Psi)^{2}+{J(0)\over 2}(\Psi^{*}\lambda^{a}\Psi)^{2}-h(\Psi^{*}\lambda^{2}\Psi), \quad \lambda^{2}\equiv S^{z}.
\end{equation}
The quadratic part reads
\begin{equation} \label{eq:3.6'}
H^{(2)}(\Psi)=H_{0}^{(2)}(\Psi)+V^{(2)}_{U}(\Psi)+V^{(2)}_{J}(\Psi),
\end{equation}
where $H_{0}^{(2)}(\Psi)$ does not include the interatomic interactions,
\begin{equation}\label{eq:3.7}
H_{0}^{(2)}(\Psi)=\sum_{{\bf p}\neq 0}\varepsilon_{\bf p}(a^{\dagger}_{\bf p}a_{\bf p})-h\sum_{{\bf p}\neq 0}(a^{\dagger}_{\bf p}\lambda^{2}a_{\bf p}).
\end{equation}
Two other terms describing the interatomic interaction have the form
\begin{equation} \label{eq:3.8}
V_{U}^{(2)}(\Psi)=U(0)\sum_{{\bf p}\neq 0}(\Psi^{*}\Psi)(a^{\dagger}_{\bf p}a_{\bf p})+{1\over 2}\sum_{{\bf p}\neq 0}U({\bf p})\left[(a^{\dagger}_{\bf p}\Psi )(\Psi^{*}a_{\bf p})+(a^{\dagger}_{\bf p}\Psi)(a^{\dagger}_{-{\bf p}}\Psi )+{\rm h.c.}\right]
\end{equation}
and
\begin{gather}
V_{J}^{(2)}(\Psi)=J(0)\sum_{{\bf p}\neq 0}(\Psi^{*}\lambda^{a}\Psi)(a^{\dagger}_{{\bf p}}\lambda^{a}a_{\bf p}) \nonumber\\
+{1\over 2}\sum_{{\bf p}\neq 0}J({\bf p})\left[(a^{\dagger}_{\bf p}\lambda^{a}\Psi)(\Psi^{*}\lambda^{a}a_{\bf p})+(a^{\dagger}_{\bf p}\lambda^{a}\Psi)(a^{\dagger}_{-{\bf p}}\lambda^{a}\Psi)+{\rm h.c.}\right], \label{eq:3.9}
\end{gather}
where we use the following notations $(\Psi^{*}\Psi)\equiv\Psi_{\alpha}^{*}\Psi_{\alpha}$, $(a^{\dagger}_{\bf p}a_{\bf p})\equiv a^{\dagger}_{{\bf p}\alpha}a_{{\bf p}\alpha}$, $(\Psi^{*}\lambda^{a}\Psi)\equiv \Psi_{\alpha}^{*}\lambda^{a}_{\alpha\beta}\Psi_{\beta}$, and so on, assuming matrix multiplication. Note that the replacement of creation and annihilation operators by $c$-numbers implies the gauge symmetry breaking and leads to non-conservation of the total number of atoms. Therefore, the problem should be considered in the grand canonical ensemble, where the chemical potential $\mu$, being a Lagrange multiplier, reflects the conservation of the total number of atoms $N=\sum_{\bf p}(a^{\dagger}_{\bf p}a_{\bf p})$. The corresponding Gibbs statistical operator for the above truncated Hamiltonian reads
\begin{equation*}\label{eq:3.10}
w(\Psi)\simeq\exp\left[\Omega-\beta\left({\cal H}^{(0)}(\Psi)+{\cal H}^{(2)}(\Psi)\right)\right],
\end{equation*}
where
\begin{gather}
{\cal H}^{(0)}(\Psi)=H^{(0)}(\Psi)-\mu{\cal V}\left(\Psi^{*}\Psi\right), \nonumber \\
{\cal H}^{(2)}(\Psi)=H^{(2)}(\Psi)-\mu\sum_{{\bf p}\neq 0}\left(a^{\dagger}_{\bf p}a_{\bf p}\right). \label{eq:3.12}
\end{gather}
The grand thermodynamic potential $\Omega$ as a function of reciprocal temperature $\beta={1/T}$, chemical potential $\mu$, and variational parameter $\Psi_{\alpha}$ is found from the normalization condition ${\rm Tr}\,w(\Psi)=1$,
\begin{equation*}\label{eq:3.13}
\Omega=\beta{\cal H}^{(0)}(\Psi)-\ln{\rm Tr}\left[ \exp(-\beta{\cal H}^{(2)}(\Psi))\right],
\end{equation*}
where the trace is taken in the space of occupation numbers of bosons with ${\bf p}\neq 0$. In the standard Bogoliubov approach, the relation between the condensate wave function and chemical potential is determined by the $c$-number part of thermodynamic potential assuming that it represents the leading term,
\begin{equation}\label{eq:3.14}
\omega^{(0)}={U(0)\over 2}(\Psi^{*}\Psi)^{2}+{J(0)\over 2}(\Psi^{*}\lambda^{a}\Psi)^{2}-h(\Psi^{*}\lambda^{2}\Psi)-\mu(\Psi^{*}\Psi),
\end{equation}
where we introduced the density of thermodynamic potential $\omega=\Omega/\beta{\cal V}$ employed when studying macroscopic dynamics of superfluid systems, both classical and relativistic \cite{JPhysA,TMF}. Up to a sign, it coincides with the pressure $P$, $\omega=-P$. The variation of Eq.~(\ref{eq:3.14}) over $\Psi^{*}_{\alpha}$ yields
\begin{equation}\label{eq:3.15}
\mu\Psi_{\alpha}-U(0)(\Psi^{*}\Psi)\Psi_{\alpha}-J(0)(\Psi^{*}\lambda^{a}\Psi)\lambda^{a}_{\alpha\beta} \Psi_{\beta}+h\lambda^{2}_{\alpha\beta}\Psi_{\beta}=0
\end{equation}
(we do not write the complex conjugate equation). This equation ensures the minimum of thermodynamic potential and gives a relation between the chemical potential and condensate wave function. The contribution of the quadratic terms in creation and annihilation operators to Eqs.~(\ref{eq:3.14}), (\ref{eq:3.15}) is examined in Refs.~\cite{Tolmachev,JPhysB} for spinless atoms.

\section{The ground state structure and excitations}

Now we use the obtained equations to study the ground state properties and corresponding single-particle excitations of a weakly interacting Bose gas of spin-1 atoms. In order to introduce the condensate density $n_{0}$, consider the normalized state vector $\zeta_{\alpha}$,
\begin{equation}\label{eq:4.1}
\Psi_{\alpha}=\sqrt{n_{0}}\zeta_{\alpha}, \quad \zeta^{*}_{\alpha}\zeta_{\alpha}\equiv(\zeta^{*}\zeta)=1.
\end{equation}
{\bf Ferromagnetic state.} In the Cartesian basis, the ferromagnetic ordering is specified by the state vector of the form \cite{Ohmi}:
\begin{equation}\label{eq:4.2}
\zeta={1\over\sqrt{2}}(1,\,i,\,0).
\end{equation}
As shown, the description of a many-body system of spin-1 constituents requires the introduction of additional parameters along with the ordinary magnetization vector. These parameters are determined by Eqs.~(\ref{eq:2.9}), (\ref{eq:2.10}). In the problem under consideration, the above ferromagnetic state vector generates the ordinary magnetization along $z$--direction,
\begin{equation}\label{eq:4.3}
\langle S^{i}\rangle=(\Psi^{*}S^{i}\Psi)=n_{0}\delta_{iz}, \quad S^{i}= (S^{x}\equiv\lambda^{7},\,S^{y}\equiv-\lambda^{5},\,S^{z}\equiv\lambda^{2})
\end{equation}
and one more parameter associated with quadrupole degrees of freedom,
\begin{equation}\label{eq:4.4}
\langle Q^{b}\rangle=(\Psi^{*}q^{b}\Psi)={n_{0}\over\sqrt{3}}\delta_{b8}.
\end{equation}
The quadrupole tensor for the ferromagnetic state becomes
\begin{equation} \label{eq:4.5}
\langle Q^{ik}\rangle=(\Psi^{*}Q^{ik}\Psi)=n_{0}\left(
  \begin{array}{ccc}
    -1/3 & 0 & 0 \\ [-1.5ex]
    0 & -1/3 & 0 \\ [-1.5ex]
    0 & 0 & 2/3 \\
  \end{array}
\right).
\end{equation}
Since $\langle Q^{xx}\rangle=\langle Q^{yy}\rangle$, the order parameter $\zeta$ is invariant with respect to rotations about $z$-axis, as it should be in the ferromagnetic state. Next, multiplying Eq.~(\ref{eq:3.15}) by $\Psi_{\alpha}^{*}$ and performing the summation over $\alpha$, one can obtain the relation between the chemical potential and condensate density for the above state vector $\zeta$:
\begin{equation} \label{eq:4.6}
\mu=n_{0}\left(U(0)+{4\over 3}J(0)\right)-h.
\end{equation}
In a similar manner, the thermodynamic potential density determined by Eq.~(\ref{eq:3.14}) is written as
$$
\omega^{(0)}={n_{0}^{2}\over 2}\left(U(0)+{4\over 3}J(0)\right)-n_{0}(h+\mu),
$$
or eliminating the condensate density by using Eq.~(\ref{eq:4.6}), one finds
\begin{equation}\label{eq:4.7}
\omega^{(0)}=-{1\over 2}{(\mu+h)^{2}\over{U(0)+(4/3)J(0)}}.
\end{equation}
In order for the equilibrium state to be stable, the thermodynamic potential density $\omega^{(0)}$ must be negative (the pressure is positive) that implies $U(0)+(4/3)J(0)>0$. 

Having defined the ferromagnetic ground state structure, we now address the issue of single-particle excitations. To obtain the corresponding spectra, let us return to the quadratic Hamiltonian given by Eqs.~(\ref{eq:3.6'})-(\ref{eq:3.9}), (\ref{eq:3.12}). Eliminating the chemical potential by using Eq.~(\ref{eq:4.6}) and taking into account the explicit form of the ground state vector and Gell-Mann matrices $\lambda^{a}$ (see Eqs.~(\ref{eq:4.1}), (\ref{eq:4.2}), (\ref{eq:A1})),  it is reduced to
\begin{equation}\label{eq:4.8}
{\cal H}^{(2)}(n_{0})={\cal H}^{(2)}_{1}(n_{0})+{\cal H}^{(2)}_{2}(n_{0}),
\end{equation}
where
\begin{equation}\label{eq:4.9}
{\cal H}^{(2)}_{1}(n_{0})=\sum_{{\bf p}\neq 0}\left[\varepsilon_{\bf p}+h+2n_{0}J({\bf p})-2n_{0}J(0)\right]a^{\dagger}_{{\bf p}z}a_{{\bf p}z},
\end{equation}
and
\begin{equation}\label{eq:4.10}
{\cal H}^{(2)}_{2}(n_{0})=\sum_{{\bf p}\neq 0}a^{\dagger}_{{\bf p}\alpha}A_{\alpha\beta}({\bf p})a_{{\bf p}\beta}+{1\over 2}\sum_{{\bf p}\neq 0}a^{\dagger}_{{\bf p}\alpha}B_{\alpha\beta}({\bf p})a^{\dagger}_{-{\bf p}\beta}+{1\over 2}\sum_{{\bf p}\neq 0}a_{{\bf p}\alpha}B^{*}_{\alpha\beta}({\bf p})a^{\dagger}_{-{\bf p}\beta}, \quad \alpha,\,\beta=x,\,y.
\end{equation}
The coefficients $A_{\alpha\beta}({\bf p})$ and $B_{\alpha\beta}({\bf p})$ form the Hermitian $A=A^{\dagger}$ and symmetric $B=B^{T}$ matrices, respectively,
\begin{equation} \label{eq:4.11}
A=\left(
  \begin{array}{ccc}
    A({\bf p}) & i{\cal A}({\bf p}) \\ [-1.0ex]
    -i{\cal A}({\bf p}) & A({\bf p}) \\
      \end{array}
\right), \quad
B=\left(
  \begin{array}{ccc}
    B({\bf p}) & iB({\bf p}) \\ [-1.0ex]
    iB({\bf p}) & -B({\bf p}) \\
      \end{array}
\right)
\end{equation}
with the following matrix elements:
\begin{gather}
A({\bf p})=\varepsilon_{\bf p}+{1\over 2}{n_{0}U({\bf p})}+{5\over 3} n_{0}J({\bf p})-n_{0}J(0)+h, \nonumber \\
{\cal A}({\bf p})=-{1\over 2}n_{0}U({\bf p})+{1\over 3}n_{0}J({\bf p})-n_{0}J(0)+h, \nonumber \\
B({\bf p})={1\over 2}n_{0}U({\bf p})+{2\over 3}n_{0}J({\bf p}). \label{eq:4.12}
\end{gather}
The first part ${\cal H}_{1}^{(2)}(n_{0})$ of the total quadratic Hamiltonian has already diagonal form with the following spectrum of single-particle excitations (see Eqs.~(\ref{eq:4.8}), (\ref{eq:4.9})):
\begin{equation}\label{eq:4.13}
\omega_{{\bf p}z}=\varepsilon_{\bf p}+h+2n_{0}\left[J({\bf p})-J(0)\right],
\end{equation}
while ${\cal H}_{2}^{(2)}(n_{0})$ should be diagonalized in creation and annihilation operators. Note that since ${\cal H}_{2}^{(2)}(n_{0})$ commutes with ${\cal H}_{1}^{(2)}(n_{0})$, it can be diagonalized independently.  To this end, we apply the Bogoliubov canonical transformation method  which allows to reduce the general Hermitian quadratic form in bosonic operators to a diagonal structure \cite{BogBog}. Therefore, let us introduce the unitary operator $U$ mixing up $a_{{\bf p}\lambda}$ and $a^{\dagger}_{{\bf p}-\lambda}$:
\begin{gather}
Ua_{{\bf p}\alpha}U^{\dagger}=\sum_{\lambda=x,y}\left[u_{\alpha\lambda}({\bf p})a_{{\bf p}\lambda}+v^{*}_{\alpha\lambda}({\bf p})a^{\dagger}_{-{\bf p}\lambda}\right], \nonumber \\
Ua^{\dagger}_{{\bf p}\alpha}U^{\dagger}=\sum_{\lambda=x,y}\left[u^{*}_{\alpha\lambda}({\bf p })a^{\dagger}_{{\bf p}\lambda}+v_{\alpha\lambda}({\bf p})a_{-{\bf p}\lambda}\right] \label{eq:4.14}
\end{gather}
and transforming ${\cal H}_{2}^{(2)}(n_{0})$ to the diagonal form
\begin{equation}\label{eq:4.15}
U{\cal H}_{2}^{(2)}(n_{0})U^{\dagger}=\sum_{{\bf p}\neq 0}\sum_{\lambda=x,y}\omega_{{\bf p}\lambda}a^{\dagger}_{{\bf p }\lambda}a_{{\bf p}\lambda}+{\cal E}_{0},
\end{equation}
where $\omega_{{\bf p}\lambda}$ are the spectra of single-particle excitations and ${\cal E}_{0}$ redefines the vacuum energy or the ground state thermodynamic potential. The creation and annihilation operators $Ua^{\dagger}_{{\bf p}\alpha}U^{\dagger}$ and $Ua_{{\bf p}\alpha}U^{\dagger}$ given by Eqs.~(\ref{eq:4.14}) must satisfy the same bosonic commutation relations as the operators $a_{{\bf p}\alpha}$ and $a^{\dagger}_{{\bf p}\alpha}$. This requirement results in the following normalization and orthogonality conditions for the functions $u_{\alpha\lambda}({\bf p})$ and $v_{\alpha\lambda}(\bf p)$:
\begin{gather*}
\sum_{\lambda=x,y}\left[u_{\alpha\lambda}({\bf p})u^{*}_{\beta\lambda}({\bf p})-v^{*}_{\alpha\lambda}({\bf p})v_{\beta\lambda}({\bf p})\right]=\delta_{\alpha\beta}, \nonumber \\
\sum_{\lambda=x,y}\left[u_{\alpha\lambda}({\bf p})v^{*}_{\beta\lambda}({\bf p})-v^{*}_{\alpha\lambda}({\bf p})u_{\beta\lambda}({\bf p})\right]=0. \label{eq:4.16}
\end{gather*}
Note that $u({\bf p})$ and $v({\bf p})$ are constructed of the same quantities as the matrices $A$ and $B$ and, therefore, they can be considered as even functions of momentum. The energies of single-particle excitations (or quasiparticles) $\omega_{{\bf p}\lambda}$ satisfy the following eigenvalue equations \cite{BogBog}:
\begin{gather*}
\sum_{\lambda=x,y}\left[A_{\alpha\lambda}({\bf p })u_{\lambda\gamma}({\bf p })+B_{\alpha\lambda}({\bf p })v_{\lambda\gamma}({\bf p})\right]=\omega_{{\bf p}\gamma}u_{\alpha\gamma}({\bf p}), \nonumber \\
\sum_{\lambda=x,y}\left[A^{*}_{\alpha\lambda}({\bf p })v_{\lambda\gamma}({\bf p })+B^{*}_{\alpha\lambda}({\bf p })u_{\lambda\gamma}({\bf p})\right]= -\omega_{{\bf p}\gamma}v_{\alpha\gamma}({\bf p}). \label{eq:4.17}
\end{gather*}
This system of homogeneous liner equations has non-zero solution when the corresponding determinant turns to zero. Therefore, taking into account Eqs.~(\ref{eq:4.11}), (\ref{eq:4.12}), one finds the equation for $\omega_{{\bf p}x}$:
$$
(A^{2}-4B^{2}-2A{\cal A}+{\cal A}^{2}-\omega_{{\bf p}x}^{2})((A+{\cal A})^{2}-\omega_{{\bf p}x}^{2})=0,
$$
which gives two different excitation spectra,
\begin{equation}\label{eq:4.18}
\omega_{{\bf p}x}^{(1)}=\varepsilon_{\bf p}+2h+2n_{0}(J({\bf p})-J(0))=\omega_{{\bf p}z}+h
\end{equation}
and
\begin{equation}\label{eq:4.19}
\omega_{{\bf p}x}^{(2)}=\left[\varepsilon_{\bf p}^{2}+2\varepsilon_{\bf p}\left(n_{0}U({\bf p})+{4\over 3}n_{0}J({\bf p })\right)\right]^{1/2}.
\end{equation}
Note that $\omega_{{\bf p}y}$ satisfies exactly the same equation as $\omega_{{\bf p}x}$ and, consequently, the corresponding spectra are identical or degenerate \cite{Ohmi}. Therefore, the ferromagnetic phase of a weakly interacting Bose gas with condensate is characterized by three types of excitations with the dispersion laws given by Eqs.~(\ref{eq:4.13}), (\ref{eq:4.18}), (\ref{eq:4.19}) and any of $\omega_{{\bf p}x}$ can be related to operators $a^{\dagger}_{{\bf p}y}$, $a_{{\bf p}y}$ in the Hamiltonian determined by Eq. (\ref{eq:4.15}) (if we take $\omega^{(1)}_{{\bf p}x}$ as $\omega_{{\bf p}y}$, then we have to take $\omega^{(2)}_{{\bf p}x}$ as $\omega_{{\bf p}x}$ and vice versa) .

The spectrum given by Eq.~(\ref{eq:4.19}) is independent of magnetic field and represents the gapless Bogoliubov mode modified by the interaction of internal degrees of freedom. At small momenta, it represents the phonon excitations,
$$
\omega_{{\bf p}x}^{\rm (II)}\approx cp, \quad c=\left[{n_{0}\over m}\left(U(0)+{4\over 3}J(0)\right)\right]^{1/2},
$$
where $c$ is a speed of sound. The requirement for the speed of sound to be real leads to the stability condition $U(0)+(4/3)J(0)>0$ obtained above.

Two other spectra $\omega_{{\bf p}z}$ and $\omega_{{\bf p}x}^{(1)}$ describe the excitations related to the internal degrees of freedom or "spin-quadrupole" waves. When the applied magnetic field is zero ($h=0$), both spectra become identical so that the system is described  by two types of single-particle excitations. Note that the quadratic form determined by Eqs.~(\ref{eq:4.10}), (\ref{eq:4.15}) must be positive definite. This requirement implies $J({\bf p})-J(0)>0$ at any momentum ${\bf p}$. It is worth stressing that in contrast to the previous studies  of spin-1 Bose-Einstein condensates \cite{Ohmi,Ho,Ueda,Kurn}, both spectra depend on the interaction parameter that is absolutely clear for the system of interacting atoms. This is due to the fact that we do not parametrize the interaction by the corresponding scattering lengths. Indeed, in doing so, $J({\bf p})=J(0)={4\pi\hbar^{2}a/m}$, where $a$ is the scattering length and the spectra become independent of the interaction parameters. Therefore, the description of the interaction effects in ultracold gases by the scattering length represents a sufficiently rough approximation that does not take into account the local character of interaction. The role of nonlocal interaction was recently discussed for ultracold Bose \cite{JPhysB,JPhysB2} and Fermi \cite{Hazlett,Caballero,Ferm-nonloc1} gases.

If the interaction Hamiltonian is SU(2) symmetric, so that it is bilinear in spin operators $S^{i}$ and does not include the quadrupole operators, then the ferromagnetic state of spin-1 condensate is characterized by the following spectra of single-particle excitations \cite{APS,PelPelSl}:
\begin{gather*}
\omega_{{\bf p}}^{(1)}=\varepsilon_{\bf p}-2J(0)n+2h, \\
\omega_{{\bf p}}^{(2)}=\varepsilon_{\bf p}+n_{0}(J({\bf p})-J(0))+h, \\
\omega_{\bf p}^{(3)}=\left[\varepsilon_{\bf p}^{2}+2\varepsilon_{\bf p}n_{0}\left(U({\bf p})+J({\bf p })\right)\right]^{1/2},
\end{gather*}
which are in agreement with other studies \cite{Ho,Ohmi,Ueda} if the interaction is taken to be of the contact type,
$$
U({\bf p})=U(0)={{g_{0}+2g_{2}}\over 3}, \quad J({\bf p})=J(0)={{g_{2}-g_{0}}\over 3},
$$
with $g_{0}$ and $g_{2}$ being related to the s-wave scattering lengths of the total spin-1 channel \cite{Ueda}. Therefore, the extension of the Hamiltonian to SU(3) symmetry leads to the appearance of new description parameters (along with the magnetization vector; see Eqs.~(\ref{eq:4.3})-(\ref{eq:4.5})) and changes the ground state and single-particle excitation spectra of ferromagnetic Bose-Einstein condensate.

{\bf Quadrupolar phase.} Equation (\ref{eq:3.15}), which ensures the minimum of thermodynamic potential, admits one more solution ($\Psi_{\alpha}=\sqrt{n_{0}}\zeta_{a}$)
\begin{equation} \label{eq:4.20}
\zeta=(0,0,1), \quad \zeta^{*}_{\alpha}\zeta_{\alpha}\equiv(\zeta^{*}\zeta)=1.
\end{equation}
In this case the relation between the condensate density and chemical potential as well as the density of thermodynamic potential are independent of the magnetic field,
\begin{equation}\label{eq:4.21}
\mu=n_{0}\left(U(0)+{4\over 3}J(0)\right)
\end{equation}
and 
\begin{equation}\label{eq:4.22}
\omega^{(0)}=-{1\over 2}{\mu^{2}\over{U(0)+(4/3)J(0)}}, \quad {U(0)+(4/3)J(0)}>0.
\end{equation}
The above state vector generates the zero magnetization,
$$
\langle S^{i}\rangle=(\Psi^{*}S^{i}\Psi)=0, \quad S^{i}= (S^{x}\equiv\lambda^{7},\,S^{y}\equiv-\lambda^{5},\,S^{z}\equiv\lambda^{2})
$$
However, at the same time, it breaks the spin-rotation symmetry,
\begin{equation}\label{eq:4.22}
\langle(S^{z})^{2}\rangle=0, \quad \langle(S^{x})^{2}\rangle=\langle(S^{y})^{2}\rangle=1
\end{equation}
showing that the spin vector fluctuates in the $x-y$ plane. Therefore, according to Eq.~(\ref{eq:4.22}) the quadrupolar state is specified by the following tensor order parameter:
$$
\langle Q_{ik}\rangle={2\over 3}\delta_{ik}-2e_{i}e_{k}, \quad {\bf e}=(0,0,\pm 1),
$$
where the unit vector ${\bf e}=(0,0,\pm1)$ or the so-called director is perpendicular to the plane of fluctuations \cite{NC}. The magnetization of this phase is zero. Next, performing the procedure of diagonalizing the corresponding quadratic Hamiltonian and employing Eqs.~(\ref{eq:4.20}), (\ref{eq:4.21}), we obtain the following three branches of the single-particle spectrum:
\begin{gather}
\omega_{{\bf p}x}=\varepsilon_{{\bf p}}+2n_{0}(J({\bf p})-J(0))+h, \nonumber \\
\omega_{{\bf p}y}=\varepsilon_{{\bf p}}+2n_{0}(J({\bf p})-J(0))-h, \nonumber \\
\omega_{{\bf p}z}=\left[\varepsilon_{\bf p}^{2}+2\varepsilon_{\bf p}\left(n_{0}U({\bf p})+{4\over 3}n_{0}J({\bf p })\right)\right]^{1/2}.  \label{eq:4.23}
\end{gather}
The comparison of Eqs.~(\ref{eq:4.7}), (\ref{eq:4.21}) for thermodynamic potentials of ferromagnetic and quadrupolar phases    allows us to conclude that the ferromagnetic phase is thermodynamically favourable.  

Finally, it is worth noting that for SU(2) symmetric Hamiltonian with competing spin and quadrupole isotropic interactions given by Eq.~(\ref{eq:2.11}), the phase diagram should exhibit a more rich structure at $T=0$. This Hamiltonian also affects the ground state thermodynamic potential, relations between the chemical potential and condensate density as well as the single-particle excitation spectra. This case deserves a separate study and lies in the field of our present interest.


\vspace{0.5cm}
\section{Conclusion}

We have proposed a general approach for describing a many-body system of interacting spin-1 atoms. It was shown that the atomic spin induces the additional description parameters related to the quadrupole degrees of freedom along with the ordinary magnetization vector. From the symmetry considerations, we found the many-body Hamiltonians of different symmetries describing the interaction of internal degrees of freedom. They include eight generators of the SU(3) group, among which there are three spin and five quadrupole operators. The SU(3) symmetric Hamiltonian was applied to study the ferromagnetic and quadrupolar phases of a weakly interacting Bose gas of spin-1 atoms with Bose-Einstein condensate. It was shown that the ferromagnetic state is thermodynamically favourable and its thermodynamic characteristics such as pressure, speed of sound, single-particle excitations are modified in comparison with those obtained from the usually employed Hamiltonian with bilinear term in spin operators. It was argued that the quadrupole degrees of freedom should be taken into account for nonlocal potentials (when parametrizing the interaction by the scattering lengths, they are inessential). Moreover, we showed that the parametrization of interaction by the scattering length may result in incomplete structure of the single-particle excitation spectra. It would be interesting to study a more rich phase diagram based on the Hamiltonian with competing spin and quadrupole isotropic interactions (see Eq.~(\ref{eq:2.11})). This problem lies in the field of our current interest. Finally, the studied Hamiltonians can be easily generalized to spin-$S$ systems by considering the generators $T^{a}$ of the SU(n) group with $n=2S+1$.

\section*{Acknowledgements}
This work was partially supported by the Ministry of Education and Science of Ukraine (State registration No. 0120U102252; Research grant with internal University No. 07-13-20). The autors thank M.~Bulakhov for useful discussions.
	


\appendix
\section{The properties of Gell-Mann matrices}

The Gell-Mann matrices, being the generators of the SU(3) group, are defined as follows:
\begin{gather}
\lambda^{1}=\left(
  \begin{array}{ccc}
    0 & 1 & 0 \\ [-2ex]
    1 & 0 & 0 \\ [-2ex]
    0 & 0 & 0 \\
  \end{array}
\right), \quad
\lambda^{2}=\left(
  \begin{array}{ccc}
    0 & -i & 0 \\[-2ex]
    i & 0 & 0 \\[-2ex]
    0 & 0 & 0 \\
  \end{array}
\right), \quad
\lambda^{3}=\left(
  \begin{array}{ccc}
    1 & 0 & 0 \\ [-2ex]
    0 & -1 & 0 \\ [-2ex]
    0 & 0 & 0 \\
  \end{array}
\right), \quad
\lambda^{4}=\left(
  \begin{array}{ccc}
    0 & 0 & 1 \\[-2ex]
    0 & 0 & 0 \\[-2ex]
    1 & 0 & 0 \\
  \end{array}
\right), \nonumber \\
\lambda^{5}=\left(
  \begin{array}{ccc}
    0 & 0 & -i \\ [-2ex]
    0 & 0 & 0 \\ [-2ex]
    i & 0 & 0 \\
  \end{array}
\right), \quad
\lambda^{6}=\left(
  \begin{array}{ccc}
    0 & 0 & 0 \\[-2ex]
    0 & 0 & 1 \\[-2ex]
    0 & 1 & 0 \\
  \end{array}
\right), \quad
\lambda^{7}=\left(
  \begin{array}{ccc}
    0 & 0 & 0 \\ [-2ex]
    0 & 0 & -i \\ [-2ex]
    0 & i & 0 \\
  \end{array}
\right), \quad
\lambda^{8}={1\over\sqrt{3}}\left(
  \begin{array}{ccc}
    1 & 0 & 0 \\[-2ex]
    0 & 1 & 0 \\[-2ex]
    0 & 0 & -2 \\
  \end{array}
\right).\label{eq:A1}
\end{gather}
They have the following property:
\begin{equation}\label{eq:A2}
{\rm Sp}\lambda^{a}\lambda^{b}=2\delta_{ab}
\end{equation}
and satisfy the commutation relations,
\begin{equation}\label{eq:A3}
[\lambda^{a},\lambda^{b}]=2if^{abc}\lambda^{c}.
\end{equation}
The structure constants $f^{abc}$ of the SU(3) group, according to Eq.~(\ref{eq:A2}), are found to be
$$
f^{abc}=-{i\over 4}{\rm Sp}\,\lambda^{c}[\lambda^{a},\lambda^{b}],
$$
whence
\begin{equation} \label{eq:A4}
f^{abc}=-f^{bac}=f^{bca}.
\end{equation}
One can easily find the numerical values of the structure constants,
\begin{equation}\label{eq:A5}
f^{123}=1, \quad f^{147}=-f_{156}=f^{246}=f^{257}=f^{345}=-f^{367}={1\over 2}, \quad f^{456}=f^{678}={\sqrt{3}\over 2}.
\end{equation}
All other numerical values of $f^{abc}$ not related to the indicated above by permutation are
zero. The anticommutator of the Gell-Mann matrices, as well as the commutator, is linear in $\lambda_{a}$:
\begin{equation} \label{eq:A6}
\{\lambda^{a},\lambda^{b}\}={4\over 3}\delta_{ab}+2d^{abc}\lambda^{c}.
\end{equation}
The coefficients $d^{abc}$, symmetric over all indices, are given by
$$
d^{abc}={1\over 4}{\rm Sp}\,\lambda^{c}\{\lambda^{a},\lambda^{b}\}.
$$
The following their values are different from zero:
\begin{gather}
d^{118}=d^{228}=d^{338}=-d^{888}={1\over\sqrt{3}}, \nonumber \\
d^{146}=d^{157}=d^{256}=d^{344}=d^{355}=-d^{247}=-d^{366}=-d^{377}={1\over 2}, \nonumber \\
d^{448}=d^{558}=d^{668}=d^{778}=-{1\over2\sqrt{3}}. \label{eq:A8}
\end{gather}
Finally, the squares of spin components $S^{x}\equiv\lambda^{7}$, $S^{y}\equiv-\lambda^{5}$, $S^{z}\equiv\lambda^{2}$ can be easily expressed in terms of the Gell-Mann matrices, 
\begin{gather}
(S^{x})^{2}={1\over 2}\left(-\lambda^{3}-{1\over\sqrt{3}}\lambda^{8}+{4\over 3}\right), \nonumber \\
(S^{y})^{2}={1\over 2}\left(\lambda^{3}-{1\over\sqrt{3}}\lambda^{8}+{4\over 3}\right), \quad
(S^{z})^{2}={1\over 3}\left(\sqrt{3}\lambda^{8}+2\right), \label{eq:A9}
\end{gather}
For the mixed products of components, we have
\begin{gather}
S^{x}S^{y}=-{1\over 2}\left(\lambda^{1}-i\lambda^{2}\right), \quad S^{y}S^{x}=-{1\over 2}\left(\lambda^{1}+i\lambda^{2}\right), \nonumber \\
S^{x}S^{z}=-{1\over 2}\left(\lambda^{4}-i\lambda^{5}\right), \quad S^{z}S^{x}=-{1\over 2}\left(\lambda^{4}+i\lambda^{5}\right), \nonumber \\
S^{y}S^{z}=-{1\over 2}\left(\lambda^{6}-i\lambda^{7}\right), \quad
S^{z}S^{y}=-{1\over 2}\left(\lambda^{6}+i\lambda^{7}\right). \label{eq:A10}
\end{gather}


\end{document}